# A Model on Genome Evolution

LiaoFu LUO

*Faculty of Physical Science and Technology, Inner Mongolia University, Hohhot 010021, China*

email: lolfcm@mail.imu.edu.cn

Abstract

A model of genome evolution is proposed. Based on three assumptions the evolutionary theory of a genome is formulated. The general law on the direction of genome evolution is given. Both the deterministic classical equation and the stochastic quantum equation are proposed. It is proved that the classical equation can be put in a form of the least action principle and the latter can be used for obtaining the quantum generalization of the evolutionary law. The wave equation and uncertainty relation for the quantum evolution are deduced logically. It is shown that the classical trajectory is a limiting case of the general quantum evolution depicted in the coarse-grained time. The observed smooth/sudden evolution is interpreted by the alternating occurrence of the classical and quantum phases. The speciation event is explained by the quantum transition in quantum phase. Fundamental constants of time dimension, the quantization constant and the evolutionary inertia, are introduced for characterizing the genome evolution. The size of minimum genome is deduced from the quantum uncertainty lower bound. The present work shows the quantum law may be more general than thought, since it plays key roles not only in atomic physics, but also in genome evolution.

Key words
genome evolution, quantum evolution, wave equation, uncertainty relation, the least action principle, evolutionary rate, evolutionary direction.

Genome is a well-defined system for studying the evolution of species. There were many publications on genome evolution. Particularly, the problem of genome size evolution has been widely discussed for a long time. The C-value enigma is still puzzling and perplexing [1][2]. On the other hand, two theoretical points, phyletic gradualism and punctuated equilibrium, were proposed to explain the macroevolution. It seems that both patterns are real facts observed in fossil evolution [3]. A deeper research question is what conditions lead to more gradual evolution and what conditions to punctuated evolution, and how to unify two patterns in a logically consistent theory. We will propose a mathematical model on genome evolution in the article. Based on a measure of diversity for DNA sequence we suggest a second-order differential equation to describe genome evolution. The directionality of the evolution is easily defined by use of the equation. Then, by putting the differential equation in a form of the least action principle, we demonstrate that the classical evolutionary trajectory can be replaced by trajectory-transitions among them in general and the concept of quantum evolution should be introduced. Thus, both classical phase (gradually and continuously) and quantum phase (abruptly and stochastically) that have been observed in the evolution can be explained in a natural way. Simultaneously, the quantum theory gives us a fully new view on the genome evolution problem.

## Basic assumptions of the model for genome evolution

**Ansatz 1**:  For any genome there exists a potential to characterize the evolutionary direction [4]

$$V(x_1,...,x_m,t) = D(x_1,...,x_m) + W_{env}(x_1,...,x_m) \qquad (1)$$

where $x_i$ means the frequency of the $i$-th nucleotide (or nucleotide pair) in DNA, $W_{env}$ is a selective potential dependent of environment, and $V$ depends on $t$ through the change of environmental variables.  $D$ means the diversity-promoting potential

$$D(x_1,...,x_m) = N \log N - \sum_i^m x_i \log x_i, \qquad N = \sum_i^m x_i \qquad (2)$$

Notice that the potential defined by (2) equals Shannon information quantity multiplied by $N$. Set $f_i = \dfrac{x_i}{N}$. We have

$$D = -N \sum f_i \log f_i \qquad (3)$$

In literatures $D$ is called diversity measure which was firstly introduced by Laxton [5]. In their studies the geographical distribution of species (the absolute frequencies of the species in different locations) was used as a source of diversity. Recently, the method was developed and applied successfully to various bioinformatics problems, for example, the intron splice site recognition[6], the promoter and transcriptional starts recognition[7], the protein structural classification[8], the nucleosome positioning prediction[9], etc. Now we shall use it to study evolutionary problem.

**Ansatz 2**:  The genome evolution equation reads as

$$\frac{d}{dt}(c^2 \frac{dx_i}{dt}) = \frac{\partial V}{\partial x_i} - f \frac{dx_i}{dt} \qquad (4)$$

where $f > 0$ is a dissipation coefficient representing the effect of fluctuation force. The parameter $c^2$ is introduced with the dimension of （time）$^2$ which represents the evolutionary inertia of the genome.

The evolutionary law can be reformulated based on Feynman's action integral [10]. Introduce a functional (called Information action)

$$S[x_1(t),...x_m(t)] = \int (\frac{c^2(t)}{2} \sum (\frac{dx_i}{dt})^2 + V(x_1,...x_m,t)) dt \qquad (5)$$

Then the solution of (4) can be expressed as

$$dS = \sum_i (\frac{\partial S}{\partial x_i} dx_i) = 0 \qquad (6)$$

as the dissipation is weak ($f$ can be neglected). Therefore, the classical evolutionary trajectory $x_i(t)$ satisfies the principle of the least action. By use of path integral quantization the evolutionary trajectory theory can be generalized to a more general quantum formalism.

**Ansatz 3**: The genome evolution obeys a general statistical law, described by the information propagator

$$U(x',t';x_0,t_0) = A\sum \exp(\frac{iS[x]}{L}) \qquad (7)$$

($x=(x_1,...x_m)$). The summation is taken over all ideal paths satisfying $x=x_0$ at $t_0$ and $x=x'$ at $t'$. Eq(7) is essentially a functional integral in mathematical term.

Here $L$ is a quantization constant of time dimension. The path integral $U(x',t';x_0,t_0)$ describes the evolution of the genomic statistical state from $t_0$ to $t'$. When $t'-t_0 > L$ the virtual variations $dx_i(t)$ may lead to $dS \gg L$ and all terms in the summation (7) would be canceled each other due to phase interference apart from those in the vicinity of classical trajectory where $S$ takes a stationary value ($dS = 0$). That is, for large $t'-t_0$ ($>L$) the classical trajectory holds. Therefore, the classical trajectory is a limiting case of the general quantum theory. However, Instead of definite trajectories, the quantum picture of trajectory transitions among them will be important during speciation if $L$ is defined by the time of the new species formation.

## Results of the model and discussions

**1  Direction of evolution**

The selective potential $W_{env}$ depends on environment. However, in stable environment both $W_{env}$ and $c^2$ are independent of $t$ and from Eq(4) one deduces

$$(V(t_1) - V(t_0)) - (K(t_1) - K(t_0)) = f \sum \int_{t_0}^{t_1} \frac{dx_i}{dt} dx_i \geq 0 \qquad (t_1 > t_0) \qquad (8)$$

where $V(t) = V(x_1(t),...,x_m(t),t)$ and

$$K(t) = \frac{c^2}{2} \sum (\frac{dx_i}{dt})^2 \qquad (9)$$

$K(t)$ is a measure of the changing rate of the nucleotide frequency. It generally takes a smaller

value as compared with the change of potential $V(t)$ for the appropriately chosen $c^2$. Thus Eq (8) means

$$D(t_1) - D(t_0) + W_{env}(t_1) - W_{env}(t_0) \geq 0 \qquad (t_1 > t_0) \qquad (10)$$

Eq (10) gives the direction of genome evolution in stable environment. The selective potential $W_{env}$ increases with $N$ for positive selection and decreases with $N$ for negative selection. For positive selection both two terms in Eq (10) are positive, namely $W_{env}(t_1) - W_{env}(t_0) \geq 0$ and $D(t_1) - D(t_0) \geq 0$. For negative selection one has $W_{env}(t_1) - W_{env}(t_0) \geq 0$ accompanying $N$ decreases and simultaneously $D(t_1) - D(t_0) < 0$. In former case the DNA sequence length increases. In latter case, although the genome size decreases the eliminated segment is often useless or even deleterious due to negative selection but the function-coding DNA is generally not deleted. In fact, DNA loss was frequently observed in genome evolution. Recent research indicated that deletional bias is a major force shaping bacterial genomes[11,12]. However, through experiments on *E coli* [13] we demonstrated that the pervasive bias towards segmental deletion is connected with the deletion of pseudogenes and other nonfunctional insertion sequences. Based on these observations it was concluded that due to natural selection, despite of the frequently occurring deletion events, the function-coding information quantity of a genome still grows in the course of evolution [4,13]. The evolutionary direction described by Eq (10) is essentially consistent with the law of function-coding information quantity growing.

## 2  Alternating occurrence of classical and quantum phases

Classical phase means the smooth evolution obeying the classical deterministic law, while the quantum phase means the sudden evolution obeying the quantum stochastic law. The present model of genome evolution predicts the alternating occurrence of both phases.

Many different estimates for the rate of evolution were made from the fossil records. As compiled by Gingerich[14][15] four hundred and nine such estimates were reported and they vary between 0 and 39 darwins in fossil linearage. Palebiological studies indicated that species usually change more rapidly during, rather than between, speciation events. The smooth evolution always occurs between speciation events and the sudden evolution preferably occurs during speciation. The former can be interpreted as the classical phase and the latter as the quantum phase in our model. Palebiological studies also indicated that the structurally more complex forms evolve faster than simpler forms and that some taxonomic groups evolve more rapidly than others. All these observations can be interpreted by the alternating occurrence of classical and quantum phases and discussed by the evolutionary equation (4) and (7).

Phyletic gradualism states that evolution has a fairly constant rate and new species arise by the gradual transformation of ancestral species. While punctuated equilibrium argues that the fossil record does not show smooth evolutionary transitions. A common pattern is for a species to appear suddenly, to persist for a period, and then to go extinct. Punctuated equilibrium states that evolution is fast at times of splitting (speciation) and comes to a halt (stasis) between splits. The theory predicts that evolution will not occur except at times of speciation. It seems that phyletic gradualism and punctuated equilibrium are contrasting and contradicting theories.[16]   However, from our model of genome evolution both smooth and sudden phase should occur in a unifying theory. By using the evolutionary equation (4) it is easily to deduce that the evolution has a range of rates, from sudden to smooth. Moreover, from the general formalism given by Eq (7) the evolutionary trajectories during speciation should be switched to quantum transitions among them . Thus, from the present theory the punctuated equilibrium and the phyletic gradualism are only the approximate description of two phases of an identical process.

## 3  Laws in classical phase

From Eq (4) and (1)(2) we deduce

$$\frac{d}{dt}(c^2 \frac{dx_i}{dt}) = -\log\frac{x_i}{N} + \frac{\partial W_{env}}{\partial x_i} - f\frac{dx_i}{dt} \tag{11}$$

Eq（11）can be written in a form of

$$\sum x_i \frac{d}{dt}(c^2 \frac{dx_i}{dt}) = D + \sum_i x_i \frac{\partial W_{env}}{\partial x_i} - \frac{f}{2}\frac{d\sum(x_i)^2}{dt} \tag{12}$$

Eqs (11) and (12) mean the Shannon information $-\log\frac{x_i}{N}$ or diversity $D$ plays a role of evolution-promoting force. Many examples show that the genome always becomes as diverse as possible and expands their own dimensionality continuously in the long term of evolution [4]. These observations are in agreement with the evolution-promoting force introduced in Eq (11). In his book "*Investigation*" Kauffman wrote："Biospheres, as a secular trend, that is, over the long term, become as diverse as possible, literally expanding the diversity of what can happen next. In other words, biospheres expand their own dimensionality as rapidly, on average, as they can." and called it "the fourth law of thermodynamics for self-constructing systems of autonomous agents"

[17]．Our proposal on evolution-promoting force is consistent with Kauffman's suggestion about the force for "expanding the diversity" in autonomous agents.

　　Segment duplication including global duplication and regional multiplication is a major force for genome evolution. The duplication can easily be deduced from Eq (11) if one assumes the selective force $\frac{\partial W_{env}}{\partial x_i} = a$ ( $a$ - a positive constant) and $c^2$ remaining a constant in certain short term. In fact, with the constant acceleration $a + \log\frac{N}{x_i} \approx a + 2$ the evolutionary trajectory obeys $x_i$；$x_{i0} + \frac{a+2}{2c^2}(t-t_0)^2$ . The time needed for $x_i$ attaining 2 $x_{i0}$ is $c\sqrt{\frac{2x_{i0}}{a+2}}$ .

Assuming $a = 2x_{i0}$ and $c = 0.01t$ or $0.1t$ （$t$ - the time of one generation）one immediately obtains the duplication time about 1% to 10% of the generation.

The driving force of genome evolution ($-\log \frac{x_i}{N}$ or $D$) and the selective force ($\frac{\partial W_{env}}{\partial x_i}$) are someway in equilibrium with the friction $f$. The selective force comes mainly from the environmental pressure and species competition; the superfluous fluctuation of the selection is represented by the friction. To be definite consider $m=4$. The force $\log \frac{N}{x_i}$ makes four kinds of nucleotide tend to equally distributed in DNA sequence ($\log \frac{N}{x_i}=2$). In stable environment an equilibrium will be established among three forces

$$\left(\frac{dx_i}{dt}\right)_{eq} = \frac{2 + \left(\frac{\partial W_{env}}{\partial x_i}\right)_\infty}{f} \qquad (13)$$

($\left(\frac{\partial W_{env}}{\partial x_i}\right)_\infty$ means the stable value of selective force). The friction coefficient $f$ is species-dependent. The larger $f$ is, the smaller the equilibrium evolutionary rate will be. For negative selection, when $\frac{\partial W_{env}}{\partial x_i}$ cancel the evolution-promoting force it leads to $\left(\frac{dx_i}{dt}\right)_{eq}=0$. This explains the evolutionary stasis of some species. For example, the study of "living fossil" lungfish showed that around 300 million years ago the lungfishes were evolving rapidly, but since about 250 to 200 million years ago their evolution has become right down. [3]. However, for a genome in suddenly-changing environment, the selective force $\frac{\partial W_{env}}{\partial x_i}$ changes rapidly. If the negative selection is strong enough then the genome cannot adapt to the sudden change of environment (for example, the food deficiency), and the species would go extinct; or new species would otherwise emerge, the genes of which can adapt to the functional needs under new environment.

**4   Basic characteristics in quantum phase**

Suppose the statistical state of the DNA evolution is represented by a wave function $y(x,t)$ that describes the probability amplitude at time $t$. The propagation of wave is determined by $U$ [10],

$$y(x,t) = \int_{-\infty}^{\infty} U(x,t;x_0,t_0) y(x_0,t_0) dx_0 \qquad (14)$$

We can prove $y(x,t)$ satisfies Schrodinger equation (see Appendix)

$$iL\frac{\partial}{\partial t}y(\text{x},t) = Hy(\text{x},t)$$

$$H = -\frac{L^2}{2c^2(t)}\sum\frac{\partial^2}{\partial x_i^2} - V(\text{x},t) \tag{15}$$

$H$ is called Hamiltonian of the genome. Eq (15) is the stochastic-generalization of the deterministic equation (4) or (11). Define

$$p_i = -iL\frac{\partial}{\partial x_i} \tag{16}$$

On account of Eq (15) one deduces

$$\int y^*(x,t)p_i y(x,t)d^m x = c^2 \frac{d}{dt}\int y^*(x,t)x_i y(x,t)d^m x \quad (d^m x = dx_1...dx_m)$$
$$\text{or} \quad \langle p_i\rangle_{av} = c^2 \langle v_i\rangle_{av} \tag{17}$$

where $v_i$ means the changing rate of nucleotide frequency. Moreover, by defining square deviations

$$\Delta x_i = \int y^*(\text{x},t)x_i^2 y(\text{x},t)d^m x - (\int y^*(\text{x},t)x_i y(\text{x},t)d^m x)^2$$
$$\Delta p_i = \int y^*(\text{x},t)(p_i)^2 y(\text{x},t)d^m x - (\int y^*(\text{x},t)p_i y(\text{x},t)d^m x)^2 \tag{18}$$

one easily obtains

$$\Delta x_i \Delta p_i \geq \frac{L^2}{4} \tag{19}$$

or

$$\sqrt{\Delta x_i \Delta v_i} \geq \frac{L}{2c^2} \tag{20}$$

for any given time. Here $\sqrt{\Delta x_i}$ means the uncertainty of nucleotide frequency and $\sqrt{\Delta v_i}$ the uncertainty of the changing rate of nucleotide frequency. Eq (19) (20) means the uncertainty relation of nucleotide frequency and its changing rate in DNA sequence. Since x and v cannot be determined simultaneously to an enough accuracy, the evolutionary trajectory cannot be accurately defined in principle.

The above discussions show that there exists good correspondence between classical evolutionary equation, Eq (4), and quantum equation, Eq (15). The quantum evolutionary equation is the logic generalization of the classical equation. The generalization is valid not only for the evolution in stable environment but also for the evolution in varying environment where the evolutionary potential $V$ and inertia $c^2$ are time-dependent.

As an application of the quantum evolutionary theory we discuss the speciation event from the view of quantum transition. Based on Schrodinger equation the speciation rate can be calculated. Suppose the initial wave function of the "old" species denoted by $y_I(\text{x})$ and the final wave function of the "new" species by $y_F(\mathrm{x})$. The transition from "old" to "new" is caused by an interaction Hamiltonian $H_{int}$ in the framework of quantum mechanics. One may assume $H_{int}$ comes from the change of evolutionary inertia, namely $H_{int} = \frac{\partial H}{\partial c}\Delta c$. Thus we have the transitional probability amplitude expressed by

$$\begin{aligned}T_{fi} &= \int y_F^*(\mathrm{x})H_{int}y_I(\mathrm{x})dx_1dx_2dx_3dx_4 \\ &= \int y_F^*(\mathrm{x})(\frac{\partial H}{\partial c}\Delta c)y_I(\mathrm{x})dx_1dx_2dx_3dx_4 \\ &= (\frac{\partial E_I}{\partial c}\Delta c)\int y_F^*(\mathrm{x})y_I(\mathrm{x})dx_1dx_2dx_3dx_4\end{aligned} \quad (21)$$

($E_I$ - the eigenvalue of $H$). Suppose

$$y_I(\mathrm{x}) = \frac{8}{pa_0}\exp(\frac{-\sum(x_i - x_{Ai})^2}{a_0})$$

$$y_F(\mathrm{x}) = \frac{8}{pa_0}\exp(\frac{-\sum(x_i - x_{Bi})^2}{a_0}) \quad (22)$$

($a_0$ - the frequency distribution width, $x_{Ai}$ and $x_{Bi}$ the frequency distribution centers for two genomes respectively). Inserting (22) into (21) one obtains

$$\int y_F^*(\mathrm{x})y_F(\mathrm{x})dx_1dx_2dx_3dx_4 = \exp(-\frac{R^2}{2a_0}), \quad R^2 = \sum(x_{Ai} - x_{Bi})^2 .$$

Therefore, the transition probability is large only for small distance $R$, since it rapidly tends to zero with increasing $\frac{R^2}{a_0}$. During speciation, corresponding to one old genome there are many candidates for the posterity with different probabilities. The most probable one is the new genome having size equal or near to the old. This explains why the observed genome-size evolution is always continuous.

## 5 Differentiation between classical and quantum phases and the observation of quantum evolution

How to differentiate classical and quantum phases in the present theory? From Eq (5) the contribution of trajectory variation $d\mathrm{x}(t)$ to $\int(\frac{c^2}{2}\sum(\frac{dx_i}{dt})^2 dt$ is proportional to $c^2$. So the variation of phase $dS/L$ in Eq (7) is much influenced by $c^2$. The large $c^2$ would make the contribution of different trajectories to the summation in (7), namely to $U(\mathrm{x}',t';\mathrm{x}_0,t_0)$, easily canceled each other, while the small $c^2$ does not. We assume that the inertia $c^2$ is a constant (denoted as $c_0^2$) for genome moving on classical trajectory under stable environment. However, it is dependent of environmental selection. During speciation the relaxed selection pressure makes the evolutionary inertia of some new species jumping to a lower value (denoted as $c_1^2$) since in this time all evolutionary events happened rapidly. If $c$ as a parameter of time dimension decreases to 1/10 of the original value or less, then the picture of classical trajectory may cease to be correct. More generally, one may assume that $c^2$ lowers from $c_0^2$ to $c_1^2 (\leq 0.01 c_0^2)$ during speciation, or lowers from $c_0^2$ to some intermediate value between $c_0^2$ and $c_1^2$ during sub-speciation. The result is the evolutionary picture switched from classical to quantum or from classical to semi-quantum upon the assumption. The point can be clarified further by looking the uncertainty relation (20). Equation (20) holds in both phases, whatever in the quantum phase or in the classical phase. However, the small $c^2$ requires the simultaneous occurrence of the larger frequency deviation and the larger deviation of the frequency changing rate. This leads to the nucleotide frequency and the frequency changing rate no longer simultaneously measurable to an enough accuracy. Therefore, the picture of classical trajectory should be replaced by a large amount of rapid trajectory-transitions. It means the quantum phase occurs and this gives a condition for new species production.

How to observe the quantum evolution in the period of new species formation? The time evolution of the genome is described by the propagation function $U(x', t'; x_0, t_0) = A \sum \exp(\frac{iS[x]}{L})$ (Eq (7)) where the summation is taken over all paths and $S$ is an integral between time $t_0$ and $t'$. The time interval $t' - t_0$ can be looked as the width of a window for observing the evolution. If $t' - t_0 > L$, only classical trajectory $x(t)$ contributes to $U(x', t'; x_0, t_0)$ due to the interference and cancelation of terms in the summation of $U$. However, if the window width $t' - t_0$ is smaller enough comparable with $L$, then the change of phase $\frac{dS[x]}{L}$ will only be in the order of one radian or less and all ideal trajectories will contribute to the propagation function. That is, the quantum transition should be observed through the small window whose width is near to or smaller than $L$. In a word, the coarse-grained evolution is observed on a classical trajectory but the quantum laws should be discovered by the fine-grained observations. Moreover, even for evolution in classical phase the quantum stochasticity can still be observed through a small enough window.

**6  Fundamental constants and minimum genome**

Two constants, quantization constant *L* and evolutionary inertia *c*, both in the dimension of time, play important roles in the quantum evolutionary theory. The former is related to the realization of the quantum picture of the evolution. The variation of the latter makes the switch from the classical phase to quantum or *vice versa*. From the estimate of the frequently observed short-term duplication the parameter *c* should be a small quantity. Set τ the average lifetime for one generation of the species. One assumes tentatively $c_0 = 10^{-1} t$ in classical phase and $c_1 = 10^{-2} t$ in quantum phase. The speciation time *L* is also dependent of τ. One assumes tentatively $L = 3 \times 10^3 \tau$ from the primary estimate of human or bacterial speciation duration.

The uncertainty relation (20) can be written in

$$\sqrt{\Delta x_i \Delta v_i} \, t \geq F$$

$$F = \frac{Lt}{2c^2}$$

*F* is the dimensionless lower bound of uncertainty relation and it is irrespective of τ. If the uncertainty $\sqrt{\Delta v_i}$ is estimated by $\frac{\sqrt{\Delta x_i}}{t}$, then the equation can be rewritten as

$$\Delta x_i \geq F \qquad (23)$$

For human the uncertainty $\sqrt{\Delta x_i}$ is estimated from the single nucleotide polymorphism, $\sqrt{\Delta x_i} = 3 \times 10^6$. Taking the above-estimated values of *L* and $c^2$ into account we prove Eq (23) is fulfilled for human. The similar prove can be done for other species.

The lower bound $F \neq 0$ means the existence of minimum genome. The present theory predicts the size of minimum genome equal $4\sqrt{F}$. Taking the estimated values $L = 3 \times 10^3 \tau$ and $c^2 = 10^{-2} t^2 - 10^{-4} t^2$ we obtain $4\sqrt{F} = 1.5 \times 10^3 - 1.5 \times 10^4$ nucleotides in accordance with the size of a typical bacteriophage genome. This is consistent with the conventional understanding on phage as species of the simplest life.

**7  Two quantum theories and two classical limits**

It is interesting to make comparison between the present quantum model for genome and the conventional quantum theory for electrons, atoms and molecules (even for some degrees of freedom of macromolecules [18]). Both systems have wave function satisfying Schrodinger equation. Both theories have their classical limit. The classical trajectory of an atom or an electron is given by the position coordinate of the particle as a function of *t*, while the classical trajectory of a genome is given by the nucleotide frequency of DNA as a function of *t*. The former is constrained by particle's energy while the latter is constrained by genome's information. Energy is conserved with time but information always grows in evolution. The atomic quantum theory has an elementary constant, the Planck's constant, while the genomic quantum theory has a corresponding constant L. The Planck's constant has dimension of (energy × **time)** but the genome constant L has dimension of **time.** The classical limit of both theories is deduced from Planck constant $h$ or genome constant $L$ approaching to zero respectively. However, the Planck's constant is universal but the constant $L$ is species-dependent because of the generation lifetimes varying largely from species to species.

# *Appendix*

Deduction of wave equation in quantum phase

Suppose the statistical state of $x=(x_1,...x_m)$ represented by a wave function $y(x,t)$.

$$y(x,e) = \int U(x,e;x_0,0) y(x_0,0) dx_0 \qquad (\varepsilon\text{- a small quantity}) \qquad (A1)$$

where $U$ is a functional integral, expressed as

$$U(x',t';x_0,t_0) = A \int \exp(\frac{iS[x]}{L}) d(x)$$
$$= \lim_{n \to \infty} A \int dx_1 ... \int dx_{n-1} \exp(\frac{iS(x_1,...x_n)}{L}) \qquad (A2)$$

and $t'-t_0 = e$ has assumed. Inserting the information action functional $S[x]$ (Eq (5))

$$S[x(t)] = \int (\frac{c^2(t)}{2}(\frac{dx}{dt})^2 + V(x,t)) dt \qquad (A3)$$

into (A2), after integrating over $(x_1,...x_{n-1})$ and taking limit $n \to \infty$ we obtain [10]

$$U(x,e;x_0,0) = (\frac{c^2(0)}{2pLie})^{m/2} \exp\frac{i}{L}\{\frac{c^2(0)(x-x_0)^2}{2e} + eV(\frac{x+x_0}{2},0)\} \qquad (A4)$$

Eq (A4) has been normalized to $U \to d(x-x_0)$ as $V=0$. In the deduction of (A4) the integral formula

$$\int_{-\infty}^{\infty} dy_1 ... \int_{-\infty}^{\infty} dy_{n-1} \exp\{-\sum_{j=0}^{n-1}\frac{1}{i}(y_{j+1}-y_j)^2\} = \frac{1}{\sqrt{n}}(ip)^{\frac{n-1}{2}} \exp\{\frac{-1}{ni}(y_n-y_0)^2\} \qquad (A5)$$

has been used. By using (A4), Eq (A1) can be written as

$$y(x,e) = (\frac{c^2(0)}{2pLie})^{m/2}) \int \exp\frac{ic^2(0)(x-x_0)^2}{2eL} \exp\{\frac{ie}{L}V(\frac{x+x_0}{2},0)\} y(x_0,0) dx_0 \qquad (A6)$$

Here $\exp\frac{ic^2(x-x_0)^2}{2eL}$ is a rapidly oscillating factor, only important near $x-x_0=0$. By setting $x_0-x=h$, Eq (A6) is rewritten as

$$y(x,e) = (\frac{c^2(0)}{2pLie})^{m/2})\int \exp\frac{ic^2(0)h^2}{2eL}\exp\{\frac{ie}{L}V(x+\frac{h}{2},0)\}y(x+h,0)dh \quad (A7)$$

After expanding $y(x+h,0)$ and $\exp\frac{ie}{L}V(x+\frac{h}{2},0)$ with respect to $h$ and $e$, finally we obtain

$$y(x,e) = y(x,0) - \frac{ie}{L}\{-\frac{L^2}{2c^2(0)}\sum\frac{\partial^2}{\partial x_i^2} - V(x,0)\}y(x,0) \quad (A8)$$

Therefore, $y(x,t)$ satisfies Schrodinger equation,

$$iL\frac{\partial}{\partial t}y(x,t) = \{-\frac{L^2}{2c^2(t)}\sum\frac{\partial^2}{\partial x_i^2} - V(x,t)\}y(x,t) \quad (A9)$$

**Acknowlegement**   The author is indebted to Dr Qi Wu for numerous helpful discussions.